\documentclass[english]{IEEEtran}
\usepackage[T1]{fontenc}
\usepackage[latin9]{inputenc}
\usepackage{float}
\usepackage{units}
\usepackage{amsmath}
\usepackage{amssymb}
\usepackage{stackrel}
\usepackage{graphicx}

\makeatletter
\newcommand{\lyxaddress}[1]{
	\par {\raggedright #1
	\vspace{1.4em}
	\noindent\par}
}

\@ifundefined{showcaptionsetup}{}{%
 \PassOptionsToPackage{caption=false}{subfig}}
\usepackage{subfig}
\makeatother

\usepackage{babel}
\begin{document}
\title{Generalization of CNOT-based Discrete Circular Quantum Walk: Simulation
and Effect of Gate Errors}
\author{Iyed Ben Slimen $^{1}$, Amor Gueddana$^{2,3}$ and Vasudevan Lakshminarayanan
$^{3,4}$}
\maketitle

\lyxaddress{{\small{}1- SysCom Lab, National Engineering School of Tunis, ENIT,
University of EL Manar, 1002 Le Belvédère, Tunis, Tunisia }}

\lyxaddress{{\small{}2- Green \& Smart Communication Systems Lab, Gres'Com, Engineering
School of Communication of Tunis, Sup'Com, University of Carthage,
Ghazela Technopark, 2083, Ariana, Tunisia. }}

\lyxaddress{{\small{}3- Theoretical \& Experimental Epistemology Lab, TEEL, School
of Optometry and Vision Science, University of Waterloo 200 University
Avenue West, Waterloo, Ontario N2l 3G1, Canada. }}

\lyxaddress{{\small{}4- Department of Physics, Department of Electrical and Computer
Engineering and Department of Systems Design Engineering , University
of Waterloo 200 University Avenue West, Waterloo, Ontario N2l 3G1,
Canada.}}
\begin{abstract}
We investigate the counterparts of random walk in universal quantum
computing and their implementation using standard quantum circuits.
Quantum walk have been recently well investigated for traversing graphs
with certain oracles. We focus our study on traversing a 1-D graph,
namely a circle, and show how to implement discrete circular quantum
walk in quantum circuits built with universal CNOT and single quit
gates. We review elementary quantum gates and circuit decomposition
and propose a a generalized version of the all CNOT based quantum
discrete circular walk. We simulated these circuits on an IBM quantum
supercomputer London IBM-Q with 5 qubits. This quantum computer has
non perfect gates based on superconducting qubits, therefore we analyze
the impact of errors on the fidelity of the Walker circuit.
\end{abstract}


\section{Introduction}

Like any other generalized tool in the field of quantum computing,
there exists a quantum version of random walk, a useful mathematical
tool to model graphs such as Markov chains. Basic concepts of quantum
walk can be found in \cite{key-1}. Quantum walk are used in computer
science and are fundamental for building quantum routers. The best
examples of quantum algorithms based on quantum walk are the searching
in an unsorted list, searching in an hypercube, element distinctness
problem, triangle problems etc ... \cite{key-2}. Circuit implementations
of quantum walk along a circle, a 2D hypercycle, a twisted toroidal
lattice graph, a complete circle and a glued tree are presented in
\cite{key-3}. Other research has addressed physical realization of
quantum walk algorithms\cite{key-4}. We distinguish two models of
quantum walk: (1). The discrete quantum walk, where we require a coin
qubit, a walker qubits and a unitary evolution operator and (2). Continuous
quantum walk, where an evolution operator is applied with no restrictions
and the time evolution of the walker is given through the Schrödinger
equation \cite{key-1}. In this paper, we address the circuit implementation
of all CNOT based Circular Quantum Discrete Random Walk (CQDRW) along
a circle and we simulate them on an IBMQ machine. In this work, we
use the London 5 qubits IBM-Q to simulate a 4 qubits CQDRW.

This paper is organized into four sections: we start section 2 by
reviewing some single quantum gates and the CNOT in order to recall
their universality. We also show how to decompose $C^{n}NOT$ into
elementary gates, specifically $C^{2}NOT$ and $C^{3}NOT$. We show
that their simulation on the IBM composer and their implementation
on London IBMQ have different outputs due to errors of the gates.
Section 3 presents the $CNOT$ based circuits for building CQDRW in
a general context. The specific simulation results for the 4 qubits
walker is presented in section 4. We focus on comparing the composer
simulation and the device execution to study the impact of errors
gates mainly due to $CNOT$s. The analysis is done by calculating
the fidelity parameter. We conclude with other possible 2-D quantum
walk implementation and some techniques to reduce the set of obtained
errors.

\section{Elementary gates for building quantum circuits}

\subsection{Universality of the CNOT and the single qubit gates}

Single qubit gates are the basic elements for building quantum circuit.
We address in this work the Identity gate $\left(I_{2}\right)$, the
Hadamard gate $\left(H\right)$, the negation gate $\left(\sigma_{x}\right)$,
the rotation by $\theta$ around $\hat{y}$ $\left(R_{y}\left(\theta\right)\right)$,
the rotation by $\alpha$ around $\hat{z}$ $\left(R_{z}\left(\alpha\right)\right)$,
the $\delta$-phase shift gate $\left(\Phi\left(\delta\right)\right)$
and the $T\left(\varphi\right)$ gate, having the following transforms
\cite{key-5,key-6,key-7}:

\begin{align}
I_{2} & =\left(\begin{array}{cc}
1 & 0\\
0 & 1
\end{array}\right)\label{eq:1}\\
H= & \frac{1}{\sqrt{2}}\left(\begin{array}{cc}
1 & 1\\
1 & -1
\end{array}\right)\label{eq:2}\\
\sigma_{x} & =\left(\begin{array}{cc}
0 & 1\\
1 & 0
\end{array}\right)\label{eq:3}
\end{align}

\begin{align}
R_{y}\left(\theta\right) & =\left(\begin{array}{cc}
cos\frac{\theta}{2} & sin\frac{\theta}{2}\\
-sin\frac{\theta}{2} & cos\frac{\theta}{2}
\end{array}\right)\label{eq:4}\\
R_{z}\left(\alpha\right) & =\left(\begin{array}{cc}
e^{i\alpha/2} & 0\\
0 & e^{-i\alpha/2}
\end{array}\right)\label{eq:5}\\
\phi\left(\delta\right) & =\left(\begin{array}{cc}
e^{i\delta} & 0\\
0 & e^{i\delta}
\end{array}\right)\label{eq:6}\\
T\left(\varphi\right) & =R_{z}\left(-\varphi\right)\phi\left(\frac{\varphi}{2}\right)=\left(\begin{array}{cc}
1 & 0\\
0 & e^{i\varphi}
\end{array}\right)\label{eq:7}
\end{align}

\begin{figure}[H]
\begin{centering}
\includegraphics[width=0.7\columnwidth]{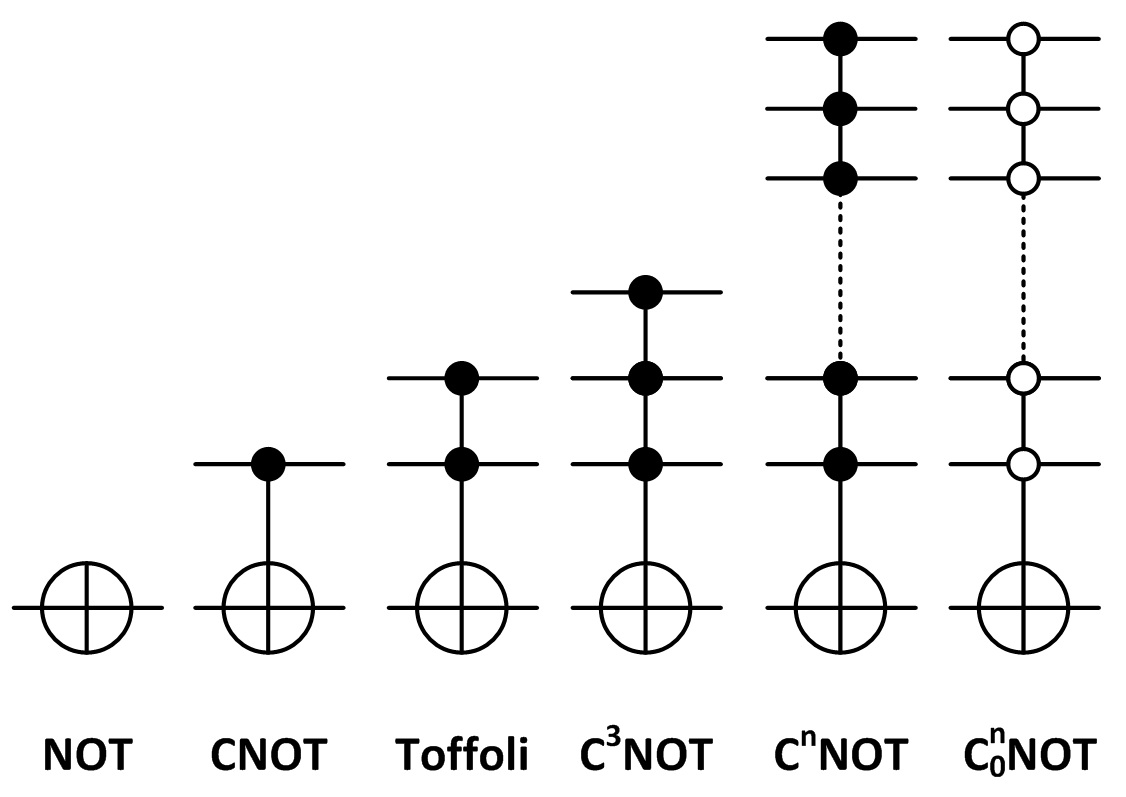}
\par\end{centering}
\caption{\label{fig:1}Controlled NOT gates}
\end{figure}

The $\sigma_{x}$ gate is also known as the NOT transform as illustrated
by figure \ref{fig:1}. The Controlled-NOT (CNOT) gate is a two qubits
gate, it performs $\sigma_{x}$ on the target qubit if and only if
the control qubit is in the state $\left|1\right\rangle $, it has
the following transform:

{\small{}
\begin{equation}
U_{CNOT}=\left(\begin{array}{cccc}
1 & 0 & 0 & 0\\
0 & 1 & 0 & 0\\
0 & 0 & 0 & 1\\
0 & 0 & 1 & 0
\end{array}\right)\label{eq:8}
\end{equation}
}{\small\par}

The set of single qubit gates and CNOT gate are universal for building
any quantum circuit. Specifically, any unitary gate acting on multiple
qubit circuit can be implemented with single qubit gates and CNOT
gates. Following paragraphs, we show how to decompose certain circuits
into a set of single qubit and CNOT gates.

The NOT controlled by two quits is known as the Toffoli gate, having
the following transform:

{\small{}
\begin{equation}
U_{Toffoli}=\left(\begin{array}{cccccccc}
1 & 0 & 0 & 0 & 0 & 0 & 0 & 0\\
0 & 1 & 0 & 0 & 0 & 0 & 0 & 0\\
0 & 0 & 1 & 0 & 0 & 0 & 0 & 0\\
0 & 0 & 0 & 1 & 0 & 0 & 0 & 0\\
0 & 0 & 0 & 0 & 1 & 0 & 0 & 0\\
0 & 0 & 0 & 0 & 0 & 1 & 0 & 0\\
0 & 0 & 0 & 0 & 0 & 0 & 0 & 1\\
0 & 0 & 0 & 0 & 0 & 0 & 1 & 0
\end{array}\right)\label{eq:9}
\end{equation}
}{\small\par}

Further generalization of the NOT gate, controlled by n qubits all
in the state $\left|1\right\rangle $, is referred as $C^{n}NOT$
gate. When the NOT gate is controlled by the n qubits all in the state
$\left|0\right\rangle $, it is denoted $C_{0}^{n}NOT$ and it's control
qubits are represented by the empty circle (figure \ref{fig:1}).

The matrix transforms of the the $C^{n}NOT$ and the $C_{0}^{n}NOT$
are obtained as follows:

{\footnotesize{}
\begin{equation}
U_{C^{n}NOT}=\left(\begin{array}{cccccc}
I_{2} & O & . & . & . & O\\
O & I_{2} &  &  &  & .\\
. &  & . &  &  & .\\
. &  &  & . &  & .\\
. &  &  &  & I_{2} & O\\
O & . & . & . & O & \sigma_{x}
\end{array}\right)=\left(\begin{array}{cc}
I_{2^{n+1}-2} & O\\
O & \sigma_{x}
\end{array}\right)\label{eq:10}
\end{equation}
}{\footnotesize\par}

{\footnotesize{}
\begin{equation}
U_{C_{0}^{n}NOT}=\left(\begin{array}{cccccc}
\sigma_{x} & O & . & . & . & O\\
O & I_{2} &  &  &  & .\\
. &  & . &  &  & .\\
. &  &  & . &  & .\\
. &  &  &  & I_{2} & O\\
O & . & . & . & O & I_{2}
\end{array}\right)=\left(\begin{array}{cc}
\sigma_{x} & O\\
O & I_{2^{n+1}-2}
\end{array}\right)\label{eq:11}
\end{equation}
}{\footnotesize\par}

where $I_{2^{n+1}-2}$ is a $\left(2^{n+1}-2\right)\times\left(2^{n+1}-2\right)$
identity matrix.

\begin{figure}[H]
\begin{centering}
\includegraphics[width=0.9\columnwidth]{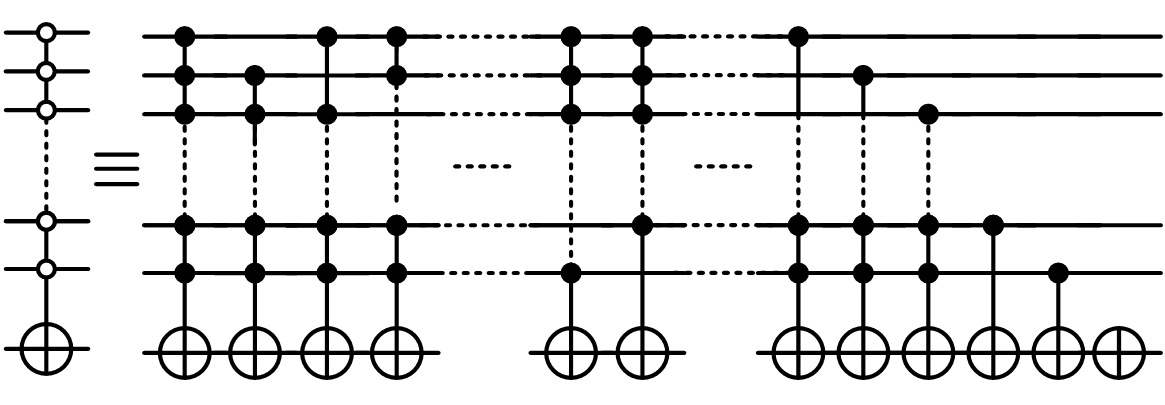}
\par\end{centering}
\caption{\label{fig2}Equivalent $C^{n}NOT$ based implementation of $C_{0}^{n}NOT$
gates}
\end{figure}

Many works have addressed circuit implementation of quantum algorithms,
such as database search algorithms, while using $C^{n}NOT$ and $C_{0}^{n}NOT$
gates \cite{key-8,key-9}, but IBMQ uses only single qubit operations
and multiple CNOTs to implement the circuits. Therefore, we show in
figure \ref{fig2} the technique used to build equivalent $C^{n}NOT$
based implementation of $C_{0}^{n}NOT$ gates \cite{key-10}.

\subsection{$CNOT$ based implementation of $C^{n}NOT$ gates}

\begin{figure}[H]
\begin{centering}
\subfloat[\label{fig:3a}$CNOT$ based implementation of Toffoli gate.]{\begin{centering}
\includegraphics[width=0.9\columnwidth]{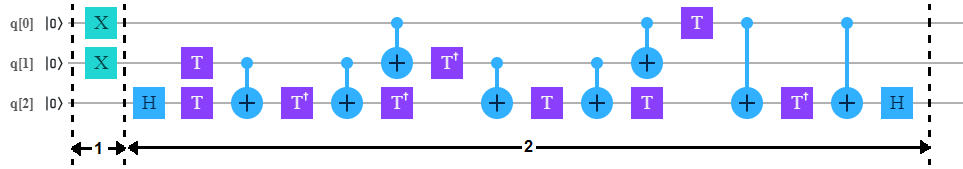}
\par\end{centering}
}
\par\end{centering}
\begin{centering}
\subfloat[\label{fig:3b}Output corresponding to the input state $\left|110\right\rangle $
after simulation on the IBMQ composer]{\begin{centering}
\includegraphics[width=0.6\columnwidth]{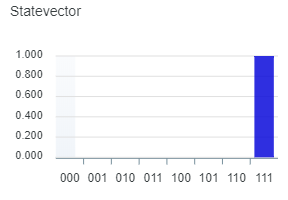}
\par\end{centering}
}
\par\end{centering}
\caption{\label{fig:3}Decomposition of the Toffoli gate and result of simulation
in the IBMQ composer for one input $\left|110\right\rangle $}
\end{figure}

Following the general decomposition method described in \cite{key-11,key-12},
we illustrate in stage 2 of figure \ref{fig:3a} the $CNOT$ based
implementation of the Toffoli gate, while the $T$ and $T^{\dagger}$
transforms refer to $T\left(\pi/4\right)$ and $T\left(-\pi/4\right)$
of equation \ref{eq:7}, respectively.

For any circuit to be simulated, IBMQ sets all input qubits automatically
to 0. In our case, the input of the decomposed Toffoli gate is set
by default to $\left|000\right\rangle $. To observe the output of
the decomposed Toffoli, we consider only the input state $\left|110\right\rangle $,
to this end, we add the two $NOT$ gates in stage 1 of figure \ref{fig:3a}
(represented by $X$ gate in IBMQ), and we illustrate the correct
output $\left|111\right\rangle $ in figure \ref{fig:3b} as obtained
after simulation on the composer.

\begin{figure}[H]
\begin{centering}
\subfloat[\label{fig:4a}Transpiled decomposition of the Toffoli gate.]{\begin{centering}
\includegraphics[width=1\columnwidth]{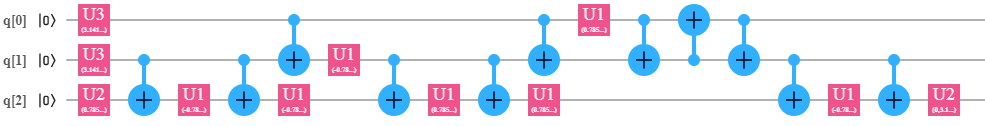}
\par\end{centering}
}
\par\end{centering}
\begin{centering}
\subfloat[\label{fig:4b}Output corresponding to the input state $\left|110\right\rangle $
after execution on the IBMQ device]{\begin{centering}
\includegraphics[width=1\columnwidth]{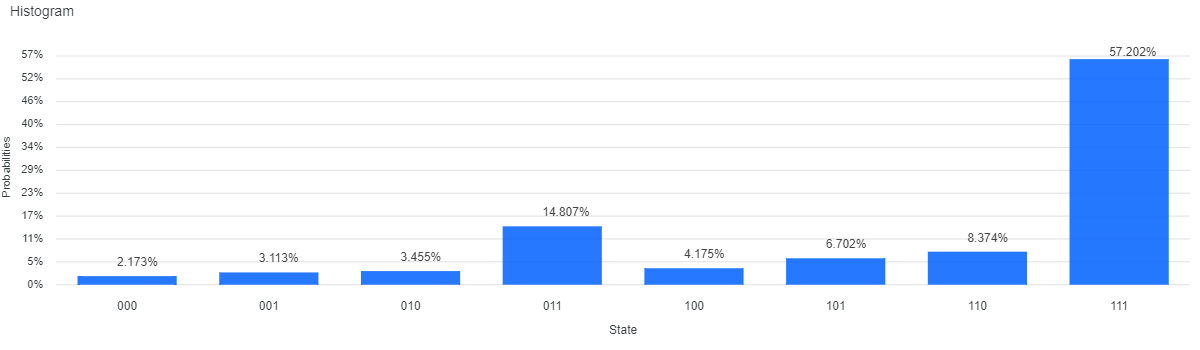}
\par\end{centering}
}
\par\end{centering}
\caption{\label{fig:4}Transpiled circuit of the $CNOT$ based implementation
of Toffoli gate.}
\end{figure}

To observe the result after execution on the real IBMQ device, a transpiled
circuit is automatically generated as given by figure \ref{fig:4a}.
The transpiled circuit performs some approximations and simplifications
based on optimizations techniques to generate transpiled circuits
that are equivalent to the original circuit. For these approximations,
all single qubits transforms given by equation \ref{eq:1} to equation
\ref{eq:7} are compiled down to physical gates based on superconducting
qubits, denoted $U_{1}$, $U_{2}$ and $U_{3}$, and given as follows:

\begin{align}
U_{1}\left(\lambda\right) & =U_{3}\left(0,0,\lambda\right)=\left(\begin{array}{cc}
1 & 0\\
0 & e^{i\lambda}
\end{array}\right)\label{eq:12}\\
U_{2}\left(\phi,\lambda\right) & =U_{3}\left(\pi/2,\phi,\lambda\right)=\frac{1}{\sqrt{2}}\left(\begin{array}{cc}
1 & -e^{i\lambda}\\
e^{i\phi} & e^{i\lambda+i\phi}
\end{array}\right)\label{eq:13}\\
U_{3}\left(\theta,\phi,\lambda\right) & =\left(\begin{array}{cc}
cos\left(\theta/2\right) & -e^{i\lambda}sin\left(\theta/2\right)\\
e^{i\phi}sin\left(\theta/2\right) & e^{i\lambda+i\phi}cos\left(\theta/2\right)
\end{array}\right)\label{eq:14}
\end{align}

From equations \ref{eq:12}, \ref{eq:13} and \ref{eq:14}, we deduce
that the two NOTs gates illustrated in stage 1 of figure \ref{fig:3a}
are implemented by $U_{3}\left(\pi,0,\pi\right)$, while the Hadamard
gate of equation \ref{eq:2} is obtained by $U_{2}\left(0,\pi\right)$.

For the same input state $\left|110\right\rangle $ applied to the
composer, we illustrate in figure \ref{fig:4b} the output after execution
on the IBMQ device. We observe a success probability of 57.202 \%
for obtaining the correct output $\left|111\right\rangle $, and various
errors for the other output states $\left|000\right\rangle $, $\left|001\right\rangle $,
$\left|010\right\rangle $, $\left|011\right\rangle $, $\left|100\right\rangle $,
$\left|101\right\rangle $ and $\left|110\right\rangle $.

\begin{figure}[H]
\begin{centering}
\subfloat[\label{fig:5a}$CNOT$ based implementation of $C^{3}NOT$ gate.]{\begin{centering}
\includegraphics[width=1\columnwidth]{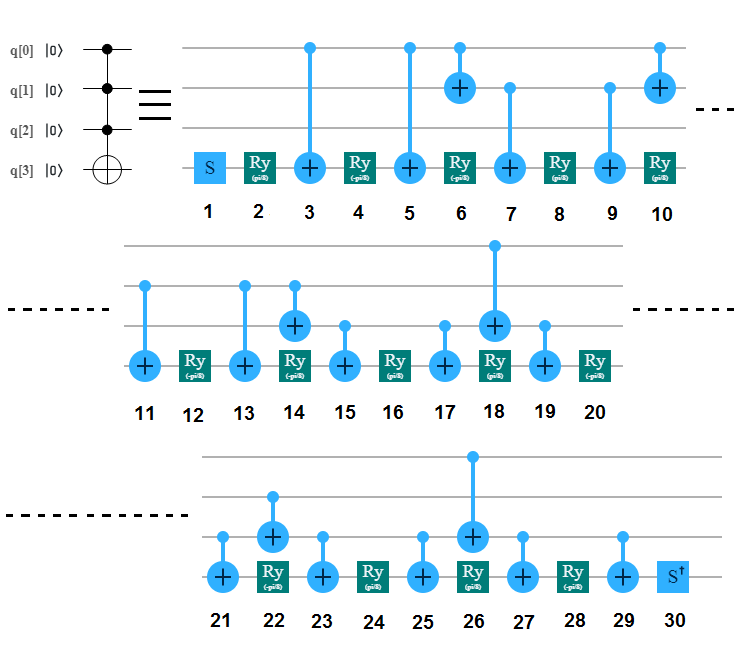}
\par\end{centering}
}
\par\end{centering}
\begin{centering}
\subfloat[\label{fig:5b}Output corresponding to the input state $\left|1110\right\rangle $
after simulation on the IBMQ composer]{\begin{centering}
\includegraphics[width=0.7\columnwidth]{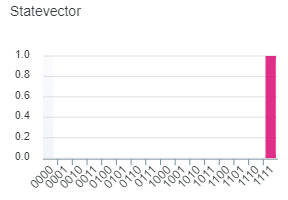}
\par\end{centering}
}
\par\end{centering}
\caption{\label{fig:5}Decomposition of the $C^{3}NOT$ gate and result of
simulation in the IBMQ composer for one input $\left|1110\right\rangle $}
\end{figure}

Following the same steps sussed to decompose the Toffoli gate \cite{key-11,key-12},
we decomposed the $C^{3}NOT$ gate and it is illustrated in figure
\ref{fig:5a}. The equivalent circuit in this figure is composed by
20 CNOTs and 16 single qubits gates. The rotation gates given in stages
2, 8, 10, 16, 18, 24 and 26 of figure \ref{fig:5a} are all identical
and equal to $R_{y}\left(\pi/8\right)$. The single qubit gates given
in stages 4, 6, 12, 14, 20, 22 and 28 of figure \ref{fig:5a} are
equal to $R_{y}\left(-\pi/8\right)$, the single qubit gate of stage
1 is $R_{z}\left(\pi/2\right)$ and it is $R_{z}\left(-\pi/2\right)$
for stage 30.

\begin{figure}[H]
\begin{centering}
\subfloat[\label{fig:6a}Output corresponding to the input state $\left|1110\right\rangle $
after execution on the IBMQ device]{\begin{centering}
\includegraphics[width=1\columnwidth]{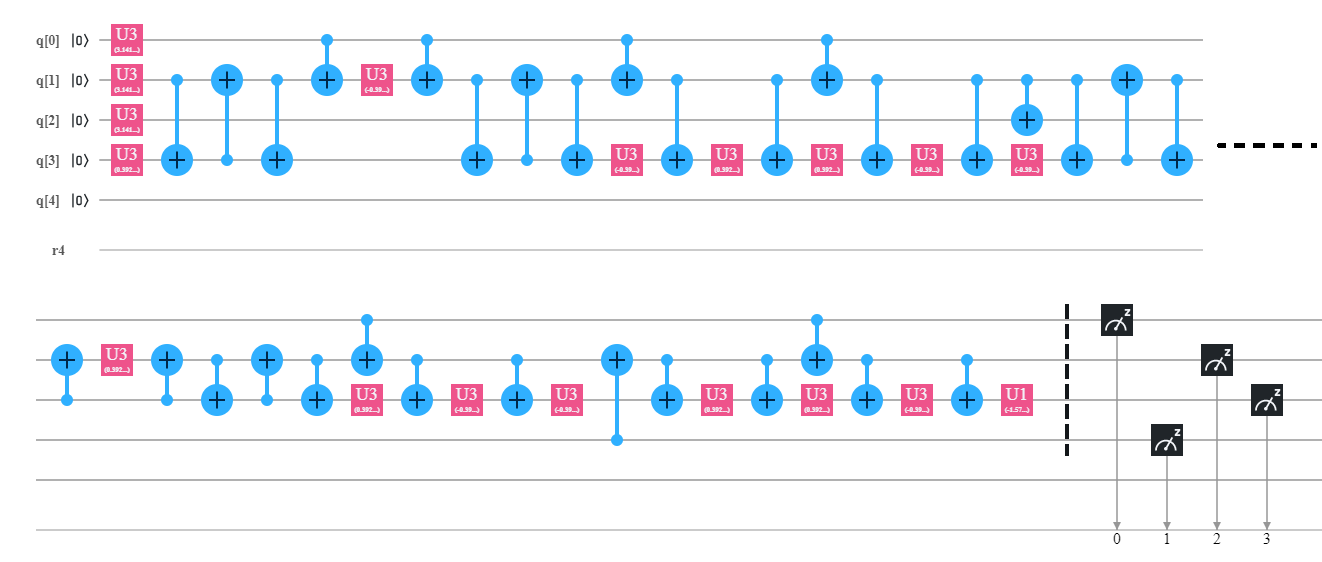}
\par\end{centering}
}
\par\end{centering}
\begin{centering}
\subfloat[\label{fig:6b}Transpiled decomposition of the $C^{3}NOT$ gate]{\begin{centering}
\includegraphics[width=1\columnwidth]{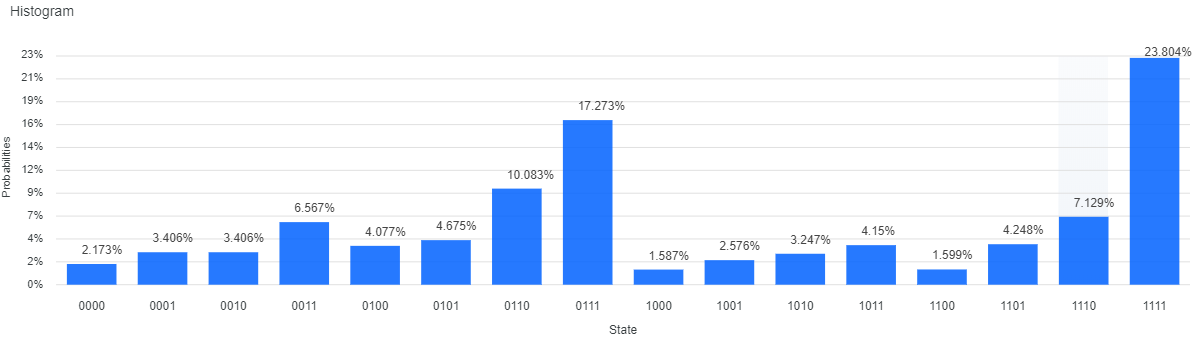}
\par\end{centering}
}
\par\end{centering}
\caption{\label{fig:6}Transpiled circuit of the $CNOT$ based implementation
of $C^{3}NOT$ gate.}
\end{figure}

For an input state $\left|1110\right\rangle $, we observe with 100
\% success the correct output $\left|1111\right\rangle $ in the composer
(figure \ref{fig:5b}). But for the according transpiled circuit (figure
\ref{fig:6b}), we observe the correct output only with 0.238 of success
probability (figure \ref{fig:6a}), this is basically due to the error
rate of the single qubit gates and the CNOT gates, which are in the
range $[3,455\times10^{-4}..\,1.058\times10^{-3}]$ and $[9.144\times10^{-3}..\,1.381\times10^{-2}]$,
respectively, according to IBMQ real device (London) \cite{key-13}.

We notice that the error of the decomposition of the $C^{3}NOT$ gate
increases exponentially depending on the number of the CNOTs used
and on the errors of all gates. We neglect in this work the errors
occurring on single qubit gates and we focus on errors due to the
CNOT. Therefore, we model these errors by an abstract probabilistic
CNOT, denoted by $A_{CNOT}^{p,\varepsilon}$, and having the following
expression \cite{key-12}:

{\small{}
\begin{eqnarray}
A_{CNOT}^{\beta,\varepsilon} & = & \left(\begin{array}{cccc}
\beta_{1} & \varepsilon_{4} & \varepsilon_{7} & \varepsilon_{10}\\
\varepsilon_{1} & \beta_{2} & \varepsilon_{8} & \varepsilon_{11}\\
\varepsilon_{2} & \varepsilon_{5} & \varepsilon_{9} & \beta_{4}\\
\varepsilon_{3} & \varepsilon_{6} & \beta_{3} & \varepsilon_{12}
\end{array}\right)\label{eq:15}
\end{eqnarray}
}{\small\par}

where $\beta=\left(\beta_{i}\right)_{1\leq i\leq4}$ represents the
probability amplitude of correctly realizing the CNOT function and
$\varepsilon=\left(\varepsilon_{j}\right)_{1\leq j\leq12}$ for $i$,
$j$$\in\mathbb{N}^{*}$, are the probability amplitudes of the errors
due to experimental realizations.

Let us highlight that the theoretical $U_{CNOT}$ of equation \ref{eq:7}
is nothing but a specific case of $A_{CNOT}^{\beta,\varepsilon}$
for all $\beta_{i}$ equal to 1 and all $\varepsilon_{j}$ equal to
0. This model of the errors is used to simulate the CQDRW in the next
section.

\section{$CNOT$-based circuits for building circular discrete quantum walker}

\begin{figure}[H]
\begin{centering}
\includegraphics[width=0.9\columnwidth]{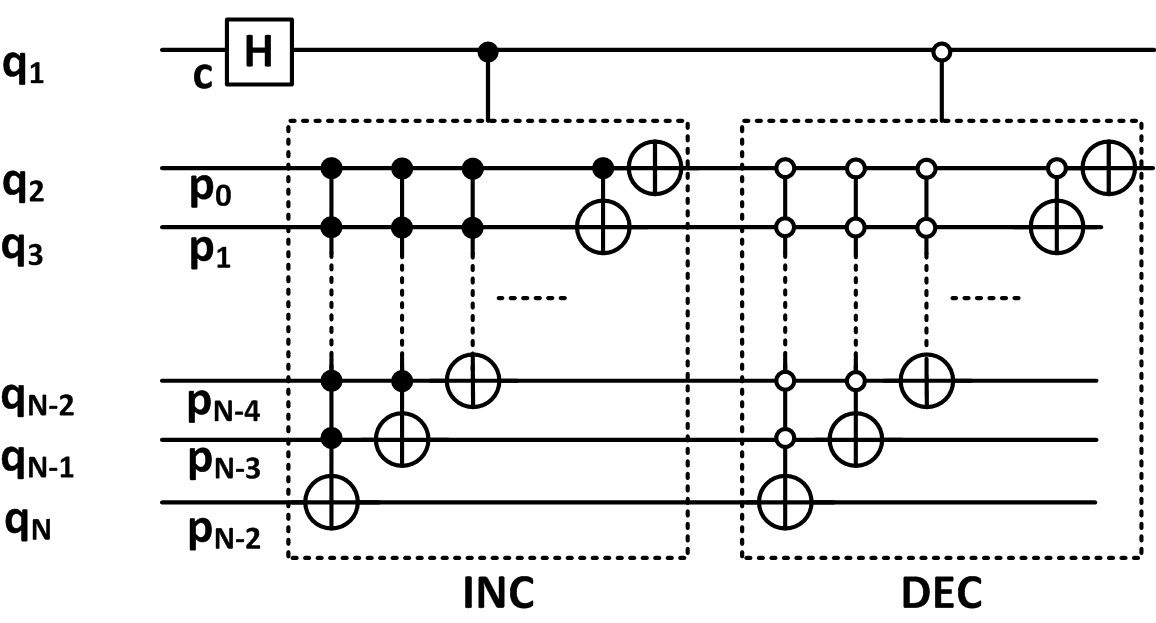}
\par\end{centering}
\caption{\label{fig:7}$C^{n}NOT$ and a $C_{0}^{n}NOT$ based implementation
of N qubits CQDRW}
\end{figure}

A CQDRW is a quantum system described in the general form by N qubits,
denoted as $\left[q_{1}..q_{N}\right]$. The qubit $q_{1}$ is used
as a coin, denoted $\left|c\right\rangle $, and N-1 qubits $\left[q_{2}..q_{N}\right]$
are used to describe a position in a circle (Figure \ref{fig:7}).

The walker could be in any position denoted $P_{k}$, for $0\leq k<2^{N-1}$,
this position is represented by the state $\left|P_{k}\right\rangle $
in the binary form as:

\begin{equation}
\left|P_{k}\right\rangle =\left|p_{N-2}^{k}\,p_{N-1}^{k}\,...\,p_{0}^{k}\right\rangle \label{eq:16}
\end{equation}

where $\left[p_{N-2}^{k}\,p_{N-1}^{k}\,...\,p_{0}^{k}\right]$ are
specific values of qubits $\left[q_{N}\,q_{N-1}\,...\,q_{2}\right]$

The N qubits system of a CQDRW at a specific position $P_{k}$, and
after performing $m$ steps, is described by the state $\left|Walker_{k}\right\rangle ^{m}$:

\begin{equation}
\left|Walker_{k}\right\rangle ^{m}=\left|P_{k}\right\rangle \otimes\left|c\right\rangle =\left|p_{N-2}^{k}\,p_{N-1}^{k}\,...\,p_{0}^{k}\right\rangle \otimes\left|c\right\rangle \label{eq:17}
\end{equation}

The walker can go one step backward or one step forward, depending
on the state of the coin, being in $\left|0\right\rangle $ or $\left|1\right\rangle $,
respectively. When the coin is in the state $\left|c\right\rangle $=
$\left|0\right\rangle $, an operator denoted $DEC$ is applied to
$\left|P_{k}\right\rangle $ and we obtain:

\begin{equation}
DEC\left|P_{k}\right\rangle =\left|P_{k-1}\right\rangle \label{eq:18}
\end{equation}

When the coin is in the state $\left|c\right\rangle $= $\left|1\right\rangle $,
an operator denoted $INC$ is applied to $\left|P_{k}\right\rangle $
and we obtain:

\begin{equation}
INC\left|P_{k}\right\rangle =\left|P_{k+1}\right\rangle \label{eq:19}
\end{equation}

Let us suppose the N qubits system with the CQDRW being at a specific
position $P_{k}$ and a coin initially at the state $\left|c\right\rangle $=
$\left|0\right\rangle $, then equation \ref{eq:17} becomes:

\begin{equation}
\begin{array}{c}
\left|Walker_{k}\right\rangle ^{0}=\left|P_{k}\right\rangle \otimes\left|0\right\rangle =\\
\left|p_{N-2}^{k}\,p_{N-1}^{k}\,...\,p_{0}^{k}\right\rangle \otimes\left|0\right\rangle 
\end{array}\label{eq:20}
\end{equation}

A single step of the walker consists of applying $H$ transform to
the coin, and then apply the appropriate $DEC$ or $INC$ operator
depending on the state of the coin, $\left|Walker_{k}\right\rangle ^{0}$
of equation \ref{eq:20} becomes $\left|Walker_{k}\right\rangle ^{1}$:

\begin{equation}
\begin{array}{c}
\left|Walker_{k}\right\rangle ^{0}\rightarrow\left|Walker_{k}\right\rangle ^{1}\\
=\frac{1}{\sqrt{2}}\left(DEC\left|P_{k}\right\rangle \otimes\left|0\right\rangle +INC\left|P_{k}\right\rangle \otimes\left|1\right\rangle \right)\\
=\frac{1}{\sqrt{2}}\left(\left|P_{k-1}\right\rangle \otimes\left|0\right\rangle +\left|P_{k+1}\right\rangle \otimes\left|1\right\rangle \right)
\end{array}\label{eq:21}
\end{equation}

A second step of the walker transforms equation \ref{eq:21} to the
following:

\begin{equation}
\begin{array}{c}
\left|Walker_{k}\right\rangle ^{1}\rightarrow\left|Walker_{k}\right\rangle ^{2}=\\
\frac{1}{\sqrt{4}}\left(\left(DEC\left|P_{k-1}\right\rangle \otimes\left|0\right\rangle +INC\left|P_{k-1}\right\rangle \otimes\left|1\right\rangle \right)\right.\\
+\left.\left(DEC\left|P_{k+1}\right\rangle \otimes\left|0\right\rangle -INC\left|P_{k+1}\right\rangle \otimes\left|1\right\rangle \right)\right)=\\
\frac{1}{\sqrt{4}}\left(\left(\left|P_{k-2}\right\rangle \otimes\left|0\right\rangle +\left|P_{k}\right\rangle \otimes\left|1\right\rangle \right)\right.\\
\left.\left(\left|P_{k}\right\rangle \otimes\left|0\right\rangle -\left|P_{k+2}\right\rangle \otimes\left|1\right\rangle \right)\right).
\end{array}\label{eq:22}
\end{equation}

According to equation \ref{eq:22}, after two steps, $\left|Walker_{k}\right\rangle ^{0}$
has walked to the positions $\left|P_{k-2}\right\rangle $, $\left|P_{k+2}\right\rangle $
and returned to initial position$\left|P_{k}\right\rangle $, with
probability amplitude equal to $\nicefrac{1}{\sqrt{4}}$, $-\nicefrac{1}{\sqrt{4}}$
and $\nicefrac{1}{\sqrt{2}}$, respectively. For $m>2$, we need to
apply each time the transform $H$ of equation \ref{eq:2}, and then
we apply the appropriate $DEC$ or $INC$ operator depending on the
state of the coin $\left|c\right\rangle $. Therefore, the general
state of the N qubits CQDRW, being initially at a position $P_{k}$
among $2^{N-1}$ positions in a circle, and after performing $m$
steps is expressed as:

\begin{equation}
\left|Walker_{k}\right\rangle ^{m}=\stackrel[i=0]{2^{m}-1}{\sum}\alpha_{i}\left|P_{i}\right\rangle \otimes\left|c\right\rangle \label{eq:23}
\end{equation}

where $\alpha_{i}$ is the probability amplitude of being in the position
$\left|P_{i}\right\rangle $ after applying the Hadamard operator
$m$ times.

A $C^{n}NOT$ and a $C_{0}^{n}NOT$ based implementation of the N
qubits walker, including $DEC$ and $INC$ possible realization is
illustrated in figure \ref{fig:7}.

Introducing all transformation rules presented in section 2 permits
us to transform the circuit of figure \ref{fig:7} into a generalized
single qubit and CNOT based implementation of any N qubits CQDRW.

\section{Simulation results}

\begin{figure}[H]
\begin{centering}
\subfloat[\label{fig:8a}$C^{3}NOT$, $C^{2}NOT$, $CNOT$,~$C_{0}^{3}NOT$,
$C_{0}^{2}NOT$ and $C_{0}NOT$ based implementation]{\begin{centering}
\includegraphics[width=0.6\columnwidth]{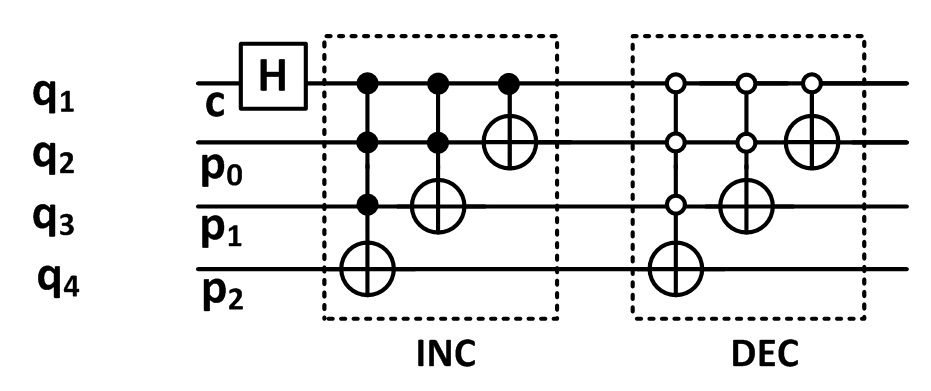}
\par\end{centering}
}
\par\end{centering}
\begin{centering}
\subfloat[\label{fig:8b}$C^{3}NOT$, $C^{2}NOT$ and $CNOT$ based implementation]{\begin{centering}
\includegraphics[width=1\columnwidth]{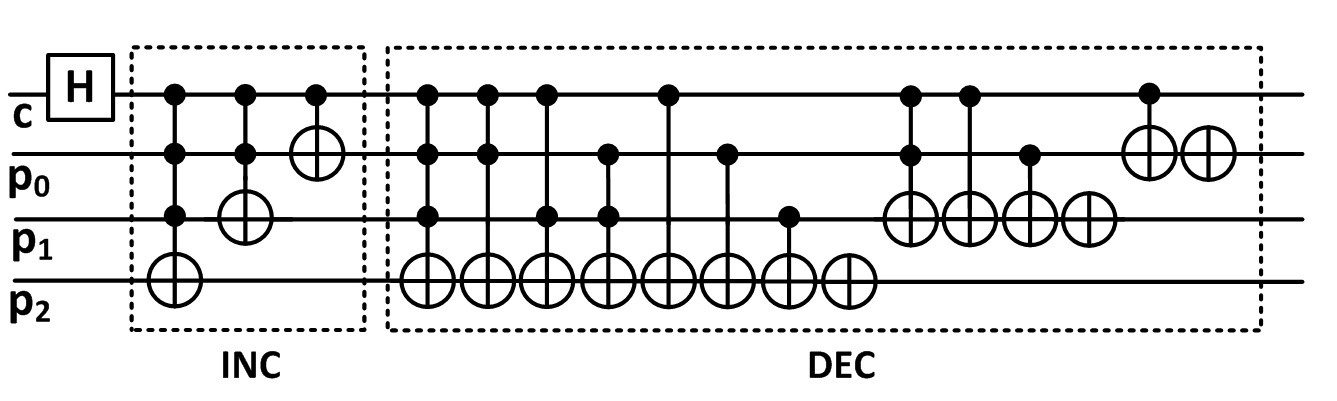}
\par\end{centering}
}
\par\end{centering}
\begin{centering}
\subfloat[\label{fig:8c}Simplified circuit version of the 4 qubits walker's
circuit]{\begin{centering}
\includegraphics[width=0.7\columnwidth]{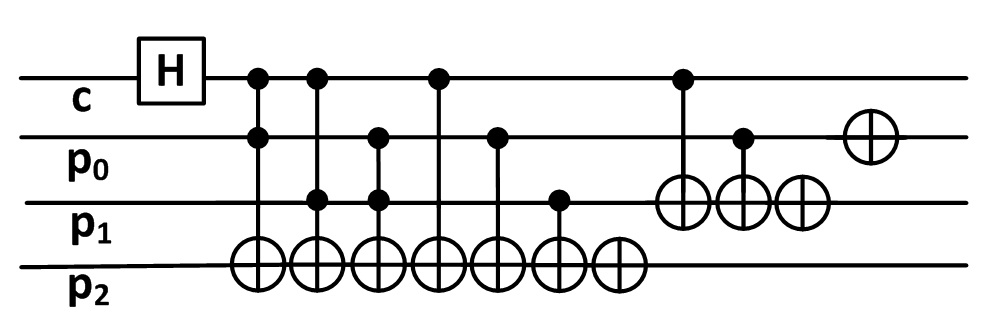}
\par\end{centering}
}
\par\end{centering}
\caption{\label{fig:8}All CNOT based circuit implementation of 4 qubits CQDRW.}
\end{figure}

In order to study the impact of the errors of the CNOT gates on the
success probability of correctly realizing the circular quantum discrete
walk, we take as an example of $N=4$ (figure \ref{fig:8a}). An equivalent
$C^{3}NOT$, $C^{2}NOT$ and $CNOT$ based implementation of figure
\ref{fig:8a} is given by figure \ref{fig:8b}. It is worth mentioning
that minimization rules detailed in \cite{key-10,key-15} permit us
to reduce the size of the circuit as given by figure \ref{fig:8c},
but since the aim of this paper is to study the impact of the errors
of the CNOT gates, we simulate the implementation of figure \ref{fig:8b},
where we introduce the decomposition of the $C^{2}NOT$ and $C^{3}NOT$
, as illustrated by figures \ref{fig:4a} and \ref{fig:5a}, respectively.

In the specific case of $N=4$, and for an initial position $P_{0}$
and a coin set to $\left|c\right\rangle =\left|0\right\rangle $,
the initial state of the 4 qubits quantum walker is expressed as:

\begin{align}
\left|Walker_{0}\right\rangle ^{0} & =\left|P_{0}\right\rangle \otimes\left|0\right\rangle =\left|p_{2}^{0}p_{1}^{0}p_{0}^{0}\right\rangle \otimes\left|0\right\rangle \label{eq:24}\\
 & =\left|000\right\rangle \otimes\left|0\right\rangle \nonumber 
\end{align}

After one step, $\left|Walker_{0}\right\rangle ^{0}$ moves to $\left|Walker_{0}\right\rangle ^{1}$as:

\begin{equation}
\begin{array}{c}
\left|Walker_{0}\right\rangle ^{0}\rightarrow\left|Walker_{0}\right\rangle ^{1}=\\
\frac{1}{\sqrt{2}}\left(\left|111\right\rangle \otimes\left|0\right\rangle +\left|001\right\rangle \otimes\left|1\right\rangle \right)
\end{array}\label{eq:25}
\end{equation}

\begin{figure}[H]
\begin{centering}
\subfloat[\label{fig:9a}Output corresponding to the input state $\left|0000\right\rangle $
after simulation on the IBMQ composer]{\begin{centering}
\includegraphics[width=0.7\columnwidth]{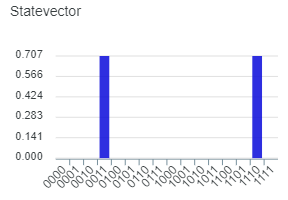}
\par\end{centering}
}
\par\end{centering}
\begin{centering}
\subfloat[\label{fig:9b}Output corresponding to the input state $\left|0000\right\rangle $
after execution on the IBMQ device]{\begin{centering}
\includegraphics[width=1\columnwidth]{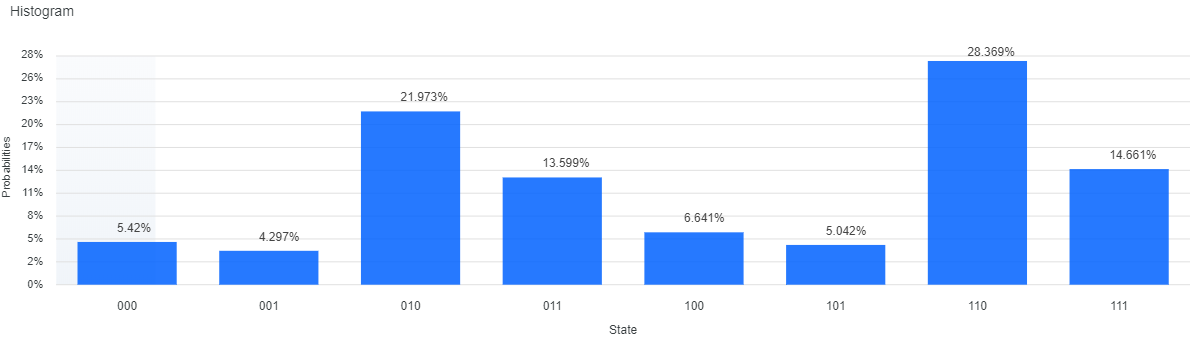}
\par\end{centering}
}
\par\end{centering}
\caption{\label{fig:9}Simulation of the 4 qubits walker on IBMQ}
\end{figure}

The state of the walker given by equation \ref{eq:25} describes exactly
the result obtained after simulation of the circuit on IBMQ (figure
\ref{fig:9a}). But on the real IBMQ device (figure \ref{fig:9b}),
$\left|Walker_{0}\right\rangle ^{0}$ of equation \ref{eq:24} becomes:

\begin{equation}
\begin{array}{c}
\left|Walker_{0}\right\rangle ^{0}\rightarrow\left|Walker_{0}\right\rangle ^{1}=\\
\sqrt{0.1466}\left|111\right\rangle \otimes\left|0\right\rangle +\sqrt{0.0311}\left|001\right\rangle \otimes\left|1\right\rangle 
\end{array}\label{eq:26}
\end{equation}

According to equation \ref{eq:26} and figure \ref{fig:9b}, we have
only 0.03 and 0.14 success probabilities for ending correctly in the
positions $\left|001\right\rangle $ and $\left|111\right\rangle $,
after just one step of the walker. To measure the performance of the
4 qubits CQDRW over all possible initial states, we refer to the fidelity
denoted $\overline{F_{Walker}}$ and given by:

{\small{}
\begin{equation}
\overline{F_{Walker}}=\left\langle \overline{\Psi_{in}|U_{Walker}^{\dagger}\rho_{t}U_{Walker}|\Psi_{in}}\right\rangle \label{eq:27}
\end{equation}
}{\small\par}

where the upper line indicates that the fidelity is obtained according
to the average over all 8 possible initial positions states{\small{}
$\left|\Psi_{in}\right\rangle =\left\{ \left|000\right\rangle \right.$,
$\left|001\right\rangle $}, {\small{}$\left|010\right\rangle $},
{\small{}$\left|011\right\rangle $}, {\small{}$\left|100\right\rangle $},
{\small{}$\left|101\right\rangle $}, {\small{}$\left|110\right\rangle $},
{\small{}$\left.\left|111\right\rangle \right\} $. $\rho_{t}$ is
given by $\rho_{t}=\left|\Psi_{out}\right\rangle \left\langle \Psi_{out}\right|$},
with {\small{}$\left|\Psi_{out}\right\rangle $} is the state at the
output of the IBMQ transpiled 4 qubits CQDRW circuit for the specific
{\small{}$\left|\Psi_{in}\right\rangle $ i}nput. The transform $U_{Walker}$
is a $16\times16$ matrix representing the ideal transform of the
4 qubits CQDRW, and obtained through Matlab simulation.

The fidelity obtained by IBMQ real device is only $17.42$ \%. This
low value is basically due to the 87 CNOTs making up the circuit (figure
\ref{fig:8b}). If we consider the error of each CNOT gate as being
equal to $1.38\times10^{-2}$ \cite{key-13}, the success probability
of each CNOT is around $0.9862$, and if we neglect the errors due
to the single qubits operations and the decoherence, the total success
probability of the entire circuit is approximately $\approx\left(0.9862\right)^{87}=29.85$
\%, which is near the fidelity value obtained in our simulation.

\begin{figure}[H]
\begin{centering}
\includegraphics[width=1\columnwidth]{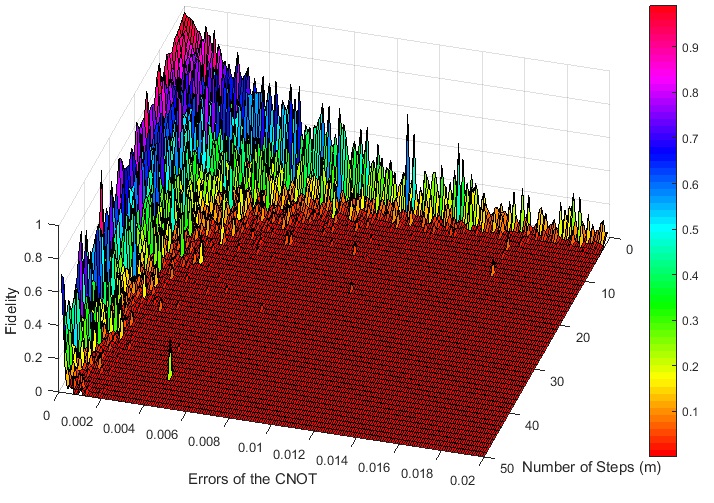}
\par\end{centering}
\caption{\label{fig:10}Fidelity of the 4 qubits CQDRW depending on the errors
of the CNOT and the number of steps}
\end{figure}

For higher number of steps $m>1$, the simulation of the 4 qubits
CQDRW would have necessitate larger circuits and a huge number of
gates. Therefore, we consider the abstract probabilistic CNOT model
of equation \ref{eq:15}, and we vary randomly all $\varepsilon=\left(\varepsilon_{j}\right)_{1\leq j\leq12}$
in a realistic range of errors $\left[10^{-5}..10^{-2}\right]$. The
results of the MATLAB stimulation of the fidelity of the walker depending
on these errors and on the number of steps $m=\left[1..50\right]$
(figure \ref{fig:10}).

According to figure \ref{fig:10}, the fidelity value of $17.42$
\% obtained by IBMQ after one step, is obtained for a CNOT error around
$10^{-2}$, which is in the error range declared by the manufacturer.
It is seen from figure \ref{fig:10} that reaching reasonable fidelity
values around 80\% or more, the error of the CNOT should be less than
$10^{-4}$, which leads us to conclude that actual state of the art
devices are still not yet ready for simulating real quantum algorithms.

\section{Conclusion}

We investigated the CQDRW both theoretically and practically and presented
a CNOT based implementation of the CQDRW in a N qubits system. We
showed through simulation on the IBMQ that the 4 qubits CQDRW system
could not exceed the fidelity value of 17 \%. We underlined the source
of the errors is related to the number of CNOT gates used in the circuit
and to the decoherence. We simulated the CQDRW for large number of
steps and showed that the error of the CNOT should be lower than $10^{-4}$
to have acceptable fidelity values. IBMQs resources constraints limited
our work to 4 qubits and informally speaking, larger CQDRW in a 5
qubits system or 2D hyper cubic quantum walks requires decomposition
of the $C^{4}NOT$ with more CNOTs, which will cause more and more
errors. Our simulation proves that working with superconducting qubits
has the major drawback of  high probability of the errors. This work
could be extended by proposing quantum error correcting codes used
to reduce the total errors of the entire circuit.

\end{document}